\documentstyle[astron,12pt,twoside,fleqn,espcrc1,epsf]{article}

\newcommand{\kms}{$\rm km\,s^{-1}$}
\newcommand{\kgs}{$\rm kg\,s^{-1}$}
\newcommand{\RJ}{$\rm \,R_J$}
\newcommand{\mum}{$\rm \mu m$}

\newcommand{\AmS}{{\protect\the\textfont2
  A\kern-.1667em\lower.5ex\hbox{M}\kern-.125emS}}

\title{
Dust en-route to Jupiter and the Galilean satellites} 

\author{Harald Kr\"uger,  Eberhard Gr\"un \\
\medskip
        Max-Planck-Institut f\"ur Kernphysik, 
        Postfach 10\,39\,80, 69029 Heidelberg, Germany}

\begin{document}
\bibliographystyle{plain}

\maketitle

\begin{abstract}

Spacecraft investigations during the last ten years have vastly 
improved our knowledge about dust in the Jovian system. 
All Galilean satellites, and probably all smaller satellites as well,
are sources of dust in the Jovian system. In-situ
measurements with the dust detectors on board the Ulysses and Galileo 
spacecraft have for the first time demonstrated the
electromagnetic interaction of charged dust grains with the 
interplanetary magnetic field and with a planetary 
magnetosphere. Jupiter's magnetosphere acts as a giant mass-velocity 
spectrometer for charged 10-nanometer dust grains. These dust
grains are released from Jupiter's 
moon Io with typical rate of $\sim 1$\kgs. 
The dust streams probe the plasma conditions in the Io plasma torus 
and can be used as a potential monitor of
Io's volcanic plume activity. The other Galilean satellites are 
surrounded by tenuous impact-generated clouds of mostly 
sub-micrometer ejecta grains. Galileo measurements
have demonstrated that impact-ejecta derived from hypervelocity
impacts onto satellites are the major -- if not the only -- constituent
of dusty planetary rings. We review the in-situ dust
measurements at Jupiter and give an update of most recent results.
\end{abstract}

\section{Introduction}

Until the 1970s, when the Pioneer~10 and Pioneer~11 spacecraft passed by 
Jupiter, the exploration of the giant planet and its satellites was 
restricted to remote astronomical observations from the Earth.
It was pure speculation whether dust would exist 
in the environment of Jupiter. Pioneer~10/11 were 
equipped with in-situ dust detectors which recorded several impacts when the 
spacecraft flew by Jupiter \cite{humes1974}. Due to the 
relatively high detection threshold of the penetration detectors, however, only 
particles larger than several micrometers could be recognized. 

The next spacecraft to visit Jupiter were Voyager~1 and Voyager~2.
Although they did not carry dedicated dust detectors on board, they 
drastically changed our knowledge of dust in the Jovian system. Jupiter's
rings were discovered by remote sensing with Voyager 1, although earlier 
hints that this faint 
dusty ring might exist came from a dip in the density of charged 
particles measured near Pioneer~11's closest approach 
to Jupiter \cite{fillius1975,acuna1976}, as well as from 
the impact events recorded by the Pioneer dust detectors.
Typical grain sizes derived from the Voyager images 
were a few micrometers for the faint gossamer ring, whereas the main ring
turned out to be composed of macroscopic rocky material. Another 
discovery by Voyager 
was tidally driven active volcanism on Io, Jupiter's 
innermost Galilean moon. At the time it was speculated that small dust grains 
entrained in Io's plumes may get accelerated away from Io by 
electromagnetic forces \cite{johnson1980,morfill1980a}. 

\begin{figure}[bt]
\vspace{-3.7cm}
\parbox{0.60\hsize}{
\epsfxsize=1.07\hsize
\vspace{8mm}
\hspace{-7mm}
\epsfbox{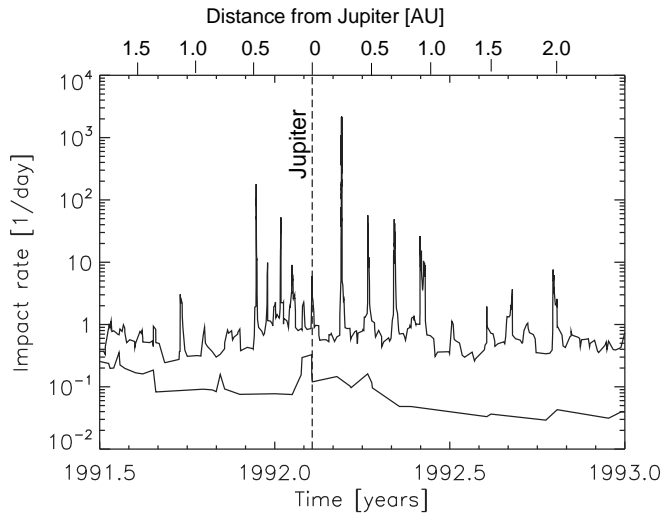}
}
\parbox{0.39\hsize}{
\vspace{-1.0cm}
        \caption{\label{ulysses_rate}
Impact rate of dust particles observed by Ulysses 
around Jupiter fly-by. The curves show all impacts recorded (upper curve) and
impacts of dust particles with masses greater than $\rm 10^{-15}\,kg$ (lower
curve). The 
impact rates are means always including six impacts. The distance from 
Jupiter is indicated at the top. Note that after Jupiter fly-by the 
craft receded from the planet at about $35^{\circ}$ jovigraphic 
latitude.
}
}
\vspace{-4.0cm}
\end{figure}

The next major step forward in the investigation of Jovian 
dust came from the Ulysses spacecraft which flew by the 
planet in 1992. Ulysses is equipped with a highly sensitive
impact-ionization dust detector capable of measuring dust grains
down to sizes of 0.1\mum\, \cite{gruen1992b}. With Ulysses, periodic 
collimated streams of dust particles with up to 2000 impacts per day
were discovered while the spacecraft was within 2\,AU from the 
giant planet \cite{gruen1993,baguhl1993} (Fig.~\ref{ulysses_rate}). 
The streams occurred at approximately monthly intervals 
($\rm 28 \pm 3$ days) and their impact directions implied that the 
grains originated from the Jovian system. 
No periodic phenomenon for small dust grains in interplanetary 
space was known before. 

Confirmation of the Jupiter dust streams came from the Galileo spacecraft
which carries a twin of the Ulysses dust detector on board 
\cite{gruen1992a}: dust ``storms'' with up to 10,000 impacts per day were 
recorded  while Galileo was within 0.5\,AU from the planet. 
\cite{gruen1996b,krueger1999a}. Since December 1995, Galileo has been the 
first man-made spacecraft in orbit about a giant planet of our solar system.
It explores Jupiter, its satellites and its huge magnetosphere.
With the Galileo dust detector the dust streams seen in interplanetary 
space were also detected within the planet's magnetosphere. The grains
showed a strong electromagnetic interaction with the Jovian magnetic field 
(see Sect.~\ref{dust_streams}). 

In December 2000 the Cassini spacecraft flew by Jupiter on its way to Saturn
and provided a unique opportunity for simultaneous two-spacecraft 
measurements of the Jovian dust streams. The Cassini
cosmic dust analyser \cite{srama2002} measured the chemical composition of 
dust stream particles in-situ for the first time.

Apart from the Jovian dust streams, Galileo allowed for studies of 
impact-generated dust clouds surrounding the
Galilean satellites \cite{krueger1999d} (Sect.~\ref{dust_clouds}), 
a tenuous dust ring in the region between 
the Galilean satellites \cite{thiessenhusen2000,krivov2002a} and further
out from the satellites \cite{krivov2002b} as well as interplanetary 
and interstellar particles captured by the Jovian magnetosphere 
\cite{colwell1998a,colwell1998b} (Sect.~\ref{dust_rings}).
The detection of most of the 
observed features was unexpected and their discovery has greatly expanded 
our knowledge about  dust in the Jovian magnetosphere.

Comprehensive reviews of more than 10 years of dust measurements with
Ulysses and Galileo focussing on Jovian dust as well as interplanetary 
and interstellar dust have also been given by Gr\"un et al. \cite{gruen2001}
and Kr\"uger et al. \cite{krueger2002a}.

\section{Jupiter dust streams}
\label{dust_streams}

\subsection{Electromagnetically interacting dust}

The impact directions of the dust stream particles measured with 
Galileo and Ulysses in interplanetary space 
were close to the 
line-of-sight direction to Jupiter. The approach direction of most streams,
however, deviated too much from the direction to Jupiter to be 
explained by gravitational forces alone. This deviation 
was correlated with the magnitude and the direction of the interplanetary 
magnetic field 
\cite{gruen1996b} which implied that  strong non-gravitational 
forces must have been acting on the grains. 
The observed 28 day period in the impact rate (Fig.~\ref{ulysses_rate}) 
was most likely caused
by changes in the tangential component of the solar wind magnetic 
field which periodically accelerated the particles towards and away 
from the ecliptic plane \cite{hamilton1993a,horanyi1993a}.
Numerical simulations showed
that only particles with velocities in excess of 
200 \kms\ and radii in the range ${\rm 5\,nm} \leq s \leq \rm 15\,nm$ 
were compatible with the observations \cite{zook1996}. Larger (smaller) 
grains did not interact enough (interacted too strongly) with 
the interplanetary magnetic field to explain the observed impact 
directions. This demonstrated that the solar wind magnetic field
acts as a giant mass-velocity spectrometer for charged dust grains.

\begin{figure}
\vspace{-2.0cm}
\parbox{0.60\hsize}{
\hspace{-1cm}
\epsfxsize=\hsize
\epsfbox{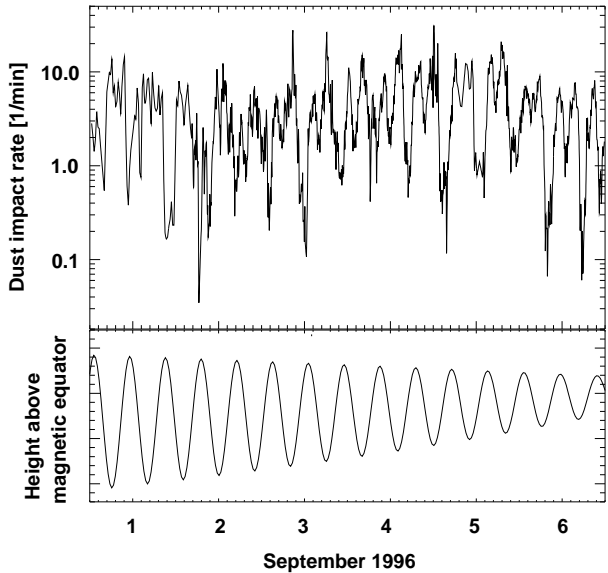}
}
\parbox{0.39\hsize}{
\vspace{-2.3cm}
        \caption{\label{g2_rate}
Top: Impact rate of Jovian dust stream particles measured in September 1996 
(Galileo's G2 orbit). In the time period shown the spacecraft approached
the planet from 60 to 10\RJ\ jovicentric distance (Jupiter radius,
\RJ\ = 71,492\,km). Note the strong fluctuations with 5 and 10~h 
periods. Bottom: Height of Galileo spacecraft above Jovian magnetic 
equator. A dipole tilted by $9.6^{\circ}$ with respect to Jupiter's 
rotation axis has been assumed. 
}
}
\vspace{-3cm}
\end{figure}

Strong electromagnetic interaction of dust grains was 
also found with the Galileo detector within the Jovian 
magnetosphere. Figure~\ref{g2_rate} shows an example of the impact 
rate measured with Galileo in the inner part of the magnetosphere.
During this and most other times when Galileo collected data in this spatial
region, the impact
rate fluctuated with 5 and 10\,h periodicities and the
fluctuations typically exceeded 2 orders of magnitude. Furthermore, the impact 
directions of the grains and the measured charge rise times and charge 
amplitudes which were used to derive particle speeds and masses showed similar 
fluctuations \cite{gruen1998}. These fluctuations were correlated with 
the position of Galileo in the Jovian magnetic field (cf. bottom panel of 
Fig.~\ref{g2_rate}). 
Due to a $9.6^{\circ}$ tilt of Jupiter's magnetic axis with respect to the planet's 
rotation axis the magnetic  equator sweeps over 
the spacecraft in either up- or downward direction every 5 h.

In addition to the 5 and 10\,h periods which are compatible with
Jupiter's rotation period, a modulation of the impact rate with Io's 
orbital period (42\,h) could also be recognized during some time intervals
(e.g. Galileo orbits E4 \cite{gruen1998}, G7 \cite{krueger1998} and C9 
\cite{krueger1999c}) while at other times an Io modulation was missing
(e.g. Galileo orbit G2, Fig.~\ref{g2_rate}). 
A detailed frequency analysis
of a two year dataset showed Io's orbital frequency as
a ``carrier frequency'' and primary source of the Jovian dust streams
\cite{graps2000}.
Jupiter's magnetic field frequency modulates Io's frequency signal,
giving rise to modulation sidelobe products seen around first order
(10~h) and harmonic (5~h) Jupiter magnetic field frequencies.
These modulation products confirm Io's role as a primary source of
the Jovian dust streams.
Io as a source can best explain the time series analysis results 
showing Io's orbit periodicity.

An Io source is also compatible with the deduced particle 
sizes of $\rm \sim 10\,nm$: photometric observations of the Io 
plumes obtained with Voyager imply a size range of 5 to 15\,nm 
\cite{collins1981}, in agreement with numerical simulations 
\cite{zook1996}. Recent Hubble Space Telescope (HST) observations
constrained the grains to be smaller than 80\,nm 
\cite{spencer1997}. Hence, given the ejection speeds of more than 
200 \kms, Io turned out to be a source for interplanetary 
and interstellar dust!

The suggested mechanism to eject dust grains from within the Jovian 
magnetosphere matched the size and velocity range of the observed stream 
particles by recognizing that these grains become positively charged in 
the Io plasma torus and can get accelerated 
by Jupiter's corotational electric field  
\cite{horanyi1993a,horanyi1993b,horanyi1997}. 
As grains traverse the various plasma regions in the torus, however, 
their charge will not remain 
constant. Dust grains escaping Io's plumes first enter the cold plasma 
torus where they become negatively charged ($\sim -3\,{\rm V}$).
Grains that reach the outer hot regions of the torus change their sign of 
charge to positive ($\sim +3\,{\rm V}$) because of secondary electron 
emission. Once positively charged, grains will be accelerated by the 
outward pointing corotational electric field. They will leave the Jovian 
system if their radii are between about 9 and 180 nm \cite{gruen1998}. 
Smaller grains remain tied to the magnetic field lines and gyrate around
them like ions do, whereas bigger grains move on gravitationally bound orbits 
which are -- depending on the particle size -- more or less affected by 
the Lorentz force. Recent investigations showed that a 
higher secondary electron yield which leads to potentials of 
$\rm -5\,V$ in the cold torus and $\rm + 5\,V$ elsewhere 
gives better agreement with the observations \cite{horanyi2000a}.

Since Io is located very close to Jupiter's equatorial plane, the particles 
are to a first order approximation accelerated outward along 
this plane. Because of the $9.6^{\circ}$ tilt of Jupiter's magnetic
field with respect to the planet's rotation axis, however, the particles also 
experience a significant out-of-plane component of the Lorentz acceleration:
particles continuously released from Io move away from Jupiter in a warped 
dust sheet which has been nick-named 'Jupiter's dusty ballerina skirt'
\cite{horanyi1993b}. A detector attached to a 
spacecraft moving in Jupiter's equatorial plane detects an increased 
number of particles when this dust sheet sweeps over its position. The 
5 and 10~h fluctuations in the dust impact rate as well as the impact 
directions of grains observed by Galileo \cite{gruen1998} can be 
explained with this scenario of electromagnetically coupled dust grains.
However, only grains within a narrow size range around 10\,nm are in agreement 
with the observed features. Smaller and larger stream particles were
not detected with the Galileo dust instrument.

The charge of a particle escaping from the Io torus strongly depends on 
variations in the plasma density and temperature in space and time 
and thus is a function of Io's position at the time of particle 
release. In fact, the position where a particle is released from the 
torus is correlated with Io's position (Graps, priv. comm.). In addition, the 
torus shows a strong dawn-to-dusk asymmetry in the plasma 
conditions that influence the escape of the dust particles.
Grain charges are more negative on the dawn side of the torus
where a lower electron temperature leads to a reduced secondary electron 
production.  Particles on the dawn side remain captured in the torus 
for longer times because of their lower positive charge. 
Six years of Galileo dust stream measurements clearly show a variation of the 
flux with Jovian local time: significantly 
higher dust fluxes were measured on the dawn and on the dusk sides
than on the noon side of Jupiter (Kr\"uger et al., in prep.) as 
predicted by numerical modelling \cite{horanyi1997}. Thus, the Jovian dust 
streams serve as tracers of the plasma conditions in the Io torus.

The fly-by of the Cassini spacecraft at Jupiter in December 2000
provided a unique opportunity for a two-spacecraft
time-of-flight measurement (Cassini-Galileo) of particles
from one collimated stream from the Jovian dust streams.
Particles in a stream were detected with Galileo as the
spacecraft was inside the Jovian magnetosphere close to the
orbit of Europa (about 12\RJ), and then
particles in the same stream were detected by Cassini
outside the magnetosphere (at 140\RJ). The Cassini
data imply that particles of different sizes 
have different phases with respect to Jupiter's rotation (Kempf et al., 
in prep.), a result
which was also seen in earlier Galileo data \cite{gruen1998}.
The comparison of the measurements
from both dust instruments, however, is hampered by the higher detection 
sensitivity of the Cassini detector with respect to 
the Galileo detector. Both instruments have detected
stream particles with different sizes and, hence, different 
phases. The analysis is ongoing and more detailed modelling to describe the
phase relation of different-sized particles is in progress. The present 
analysis indicates particle speeds of about $400 \,\mathrm{km\,s}^{-1}$.
This value is in agreement with speeds for 5~nm particles as
derived from dynamical modelling 
and earlier studies
of the Jovian dust stream dynamics \cite{zook1996}.

The Cassini dust instrument is equipped with a time-of-flight 
mass spectrometer
which measures the elemental composition of dust grains with 
a mass resolution $\rm M/\Delta M \approx 100$. During Cassini's 
approach to Jupiter impact spectra of a few hundred dust stream
particles have been measured and their chemical composition 
reflects the chemistry found on Io. With the Cassini instrument the
surface composition of a satellite other than our Moon has been measured 
directly.

\begin{figure}[bt]
\vspace{-0.3cm}
\hspace{2.5cm}
\epsfxsize=.6\hsize
\epsfbox{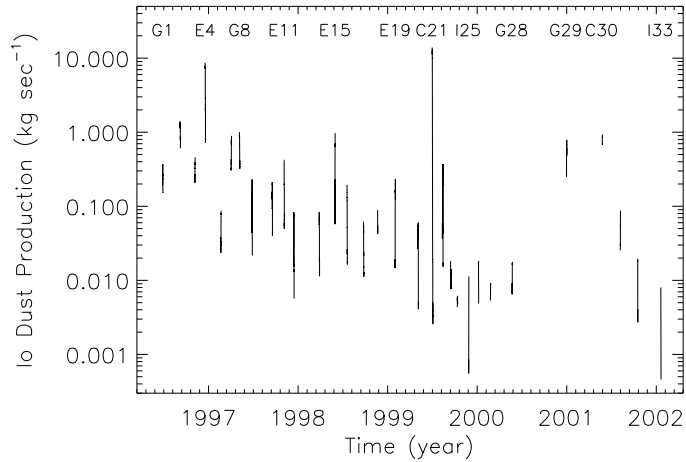}
\vspace{-1.0cm}
        \caption{\label{ioflux}
Total dust production rate of Io assuming that the grains are ejected into a 
cone with $35^{\circ}$ opening angle centered at Jupiter. 
Each vertical bar represents 
data from one Galileo orbit. The height of the bar shows the 
dust production rate derived from measurements between 10 to 30 \RJ\
from Jupiter. The data have been corrected for a Jovian local time 
variation of the dust emission from the Io torus and for a long-term 
change of the dust instrument sensitivity (Kr\"uger et al. in prep.).
The labels of individual Galileo orbits are indicated at the top.
No dust stream measurements were collected during Galileo orbits 5 and 13.
}
\end{figure} 

\subsection{Io as a source of dust in the Jovian system}

How significant is Io as a source of cosmic dust? How does the amount of
dust ejected compare with other dust sources in the solar system? 
With a simple 
calculation we can derive the total dust production rate of Io. Given 
the spread of Io dust along and away from Jupiter's equatorial 
plane, we assume a cone-shaped emission pattern of dust originating
at Jupiter. We assume a cone opening angle of $35^{\circ}$ and 
isotropic dust emission towards all jovigraphic longitudes. Although
Galileo measurements were obtained only along the Jovian equatorial 
plane, this opening angle is justified by the Ulysses measurements.
Ulysses measured the dust streams at 
$35^{\circ}$ jovigraphic latitude after Jupiter fly-by (cf. 
Fig.~\ref{ulysses_rate}). For a given impact rate $R$, particle density 
$\rho = 2\,{\rm g\,cm^{-3}}$, particle radius $a = 10\,{\rm nm}$, 
a detector sensitive area of $A = 0.02\,{\rm m^2}$ and a cone
radius $r = $ 30\RJ\, the total amount of dust emitted from Io per 
second can be calculated.
With $R = 0.1 \ldots 100$ impacts per minute detected from 1996 to 2001 the
average dust ejection rate is $\rm 10\,g\,s^{-1}$ to 10 \kgs (Fig.~\ref{ioflux}).
If we take a typical value of 1\kgs\ of dust
and compare it with $\rm 1\,ton\,sec^{-1}$ of plasma ejected from Io into the 
torus, the dust amounts to only 0.1\% of the total mass released.
These numbers indicate that Io's volcanic plumes are also a minor 
source for interplanetary dust compared with comets or main belt 
asteroids \cite{whipple1987}.  
Io, however, turns out to be a major dust source for the Jovian system
itself. The total mass of dust produced by Io as 10 nm-sized 
particles is comparable to the mass of dust ejected as micrometer-sized
particles by the other Galilean satellites, which have no
volcanic activity (Sect.~\ref{dust_clouds}).

The Jovian dust stream measurements can serve as a monitor of Io's volcanic 
plume activity. With Galileo imaging ten active plumes have been 
observed which is comparable with nine plumes 
seen by Voyager \cite{mcewan1998}. At least tow types of plumes can 
be distinguished: large, faint ones, with short-lived or intermittent 
activity (Pele-type) or small, bright, long-lived ones (Prometheus-type). 
The most powerful plume ever detected on Io, Pele, is the archetype of 
the first category and
was observed to an altitude of more that 400\,km \cite{spencer1997}. 
Pele is also the location of the most stable high-temperature hot-spot on 
Io and is probably related to an active lava lake. Plumes are normally 
related to hot spots but not vice versa.
The Pele plume is known 
to be rich in $\rm S_2$ gas as well as $\rm SO_2$ \cite{spencer2000}.
Although it has been suggested that the Pele plume may be a pure gas
plume, plume observations can also be interpreted as due to 
very fine ($ \rm \leq 80\,nm$) particulates \cite{spencer1997}.

It is of special interest to see whether variations in the 
dust production rate deduced from the dust stream measurements 
can be related to the activity of the Pele or other plumes on Io, or 
to the total thermal output of the satellite. A correlation with the
activity of the Pele plume seems most promising 
because only the most powerful plumes are expected to accelerate the grains
to sufficient altitudes so that they can finally escape from
the satellite \cite{johnson1980,ip1996}.

The dust production of Io for individual orbits of Galileo is 
shown in Fig.~\ref{ioflux}. Here, the vertical bars indicate the variation
in the derived dust production rate if we vary the jovicentric distance at 
which the dust flux is taken between 
10 and 30\RJ\ during one orbit. 
This reveals a strong variation in the dust production 
rate from orbit to orbit which is up to two orders of magnitude. If 
the plasma conditions in the Io torus and the Jovian magnetic field  did not 
change too drastically from orbit to orbit, it reflects the variation 
of the activity of the Io plumes. 

We have compared the dust production rate shown in Fig.~\ref{ioflux}
with the total thermal output of Io deduced from Galileo near-infrared 
measurements (Spencer, priv. comm.). Unfortunately, this did not 
give a clear picture. This negative result, however, is not too surprising
because Io's overall thermal output is not very well correlated with 
plume activity.  The Pele plume was observed  in July 1995, July 1996,
 December 1996 and 
possibly July 1997 \cite{mcewan1998}. It was absent in June 1996, 
February 1997, June 1997 and July 1999. Although, the strong drop in 
the dust impact rate from 
December 1996 to February 1997 (E4 to E6 orbit) is consistent with 
these detections/non-detections, for other measurements it is not.
Especially, the non-detection of the plume on 2 July 1999 
is in contradiction with the large measured dust emission.

A correlation of the in-situ dust measurements with either Galileo
or Earth-based imaging observations turns out to be very difficult because 
the imaging observations represent only sporadic glimpses.
Many more observations would be needed to establish a firm link between
the Galileo dust measurements and the activity of (an) individual 
plume(s) on Io. The picture is further complicated by the fact that the 
plume activity sometimes changes on timescales of days to weeks. 
Ideally,
one would need imaging observations at exactly the same time as the
dust measurements.

We have also estimated the Io dust production from the 
measurements of Galileo and Ulysses in interplanetary space out
to 1\,AU from Jupiter assuming again that the dust is uniformly
distributed into a cone of $35^{\circ}$. This leads to unrealistically
high dust production rates of more than $10^7$ \kgs. It indicates 
that this simple picture cannot be extrapolated to interplanetary 
space and that the dust is not distributed uniformly  to such large
distances. Rather, the dust particle trajectories must undergo 
some focussing effect due to electromagnetic interaction with the
interplanetary magnetic field.


\begin{figure}[bt]
\vspace{-0.3cm}
\hspace{2.5cm}
\epsfxsize=.6\hsize
\epsfbox{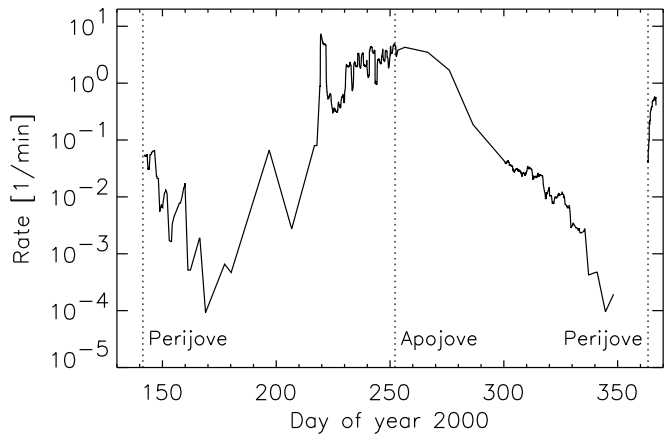}
\vspace{-0.7cm}
        \caption{\label{g28_rate}
Impact rate of dust stream particles measured with Galileo between
May and December 2000 (Galileo G28 orbit). Galileo perijove (at
$\sim 8$\RJ\ distance from Jupiter) and apojove ($\sim 280$\RJ)
are indicated. The impact rate has been averaged over a 40~h time
interval. Due to different modes of dust instrument read-out 
the data have a time resolution varying from less than an hour to 
more than 20 days (28). The dust streams were 
outside the field of view of the
dust detector before day 141 and again between day 350 and 363.
}
\end{figure}
\nocite{krueger2001a}

Additional evidence for such a focussing effect came from Galileo 
measurements in 2000 when the spacecraft has 
left the Jovian magnetosphere for the first time since 1995. 
Measurements outside the magnetosphere at 
a distance of $\sim 280$\RJ\ (0.13~AU) from Jupiter 
gave a surprisingly high impact rate of up to 10 impacts per minute
(Fig.~\ref{g28_rate}). 
This value was comparable with the rates detected both in 
interplanetary space (Fig.~\ref{ulysses_rate}) and close to Jupiter 
during Galileo's early orbital mission (Fig.~\ref{g2_rate}). 

In May and June 2000 (days 145 to 170), while Galileo was receding 
from Jupiter (from 10 to 170\RJ), the impact rate dropped by more 
than two orders of
magnitude (from 0.05 to 0.0005 impacts per minute). This drop was
close to the inverse square of the source distance. 
When Galileo was outside the magnetosphere, beyond $\sim\,$200\RJ\ 
from Jupiter (after day 180), the impact rate increased by about four 
orders of magnitude.
Between August and October 2000 (days 230 to 280), Galileo remained more 
or less stationary with respect to Jupiter and Io, and the 
impact rate remained remarkably constant for about two months with 
roughly 1 impact per minute. Assuming -- as before -- 
that dust particles get ejected into a $35^{\circ}$ cone, this leads 
to a dust production of Io of $\sim 100\,$\kgs. It seems unlikely that
such a high dust production is maintained over such 
a long time period.
More likely is a focussing effect of the grains 
due to the boundary between the Jovian magnetosphere and the interaction
with the interplanetary magnetic field. 
Interestingly, the impact directions measured with Galileo indicate that 
the grains approached the sensor from a direction very close to the 
ecliptic plane. 
Similarly high impact rates were was also detected with the Cassini 
dust instrument \cite{srama2002} in September 2000 at $\sim 0.3$AU
from Jupiter when the spacecraft was approaching the planet
(Kempf et al., in prep.).

Frequency analysis of the Galileo dust impact rates measured beyond 
$\sim 250$ \RJ\ did not reveal 5 and 10\, h periodicities as was seen 
within the magnetosphere. Instead, a strong peak at Io's orbital 
period showed up in the frequency spectrum (A. Graps, priv. comm.), 
much stronger than seen close to Jupiter.

\section{Dust-enshrouded satellites}

\label{dust_clouds}

Between December 1995 and January 2002 the Galileo
spacecraft had a total of 31 targeted encounters with all four
Galilean satellites. During many of these fly-bys the 
impact rate of dust grains showed a sharp peak within
about half an hour centered on closest approach to the satellite
\cite{gruen1997b,gruen1998,krueger1998}. 
This indicated the existence of dust concentrations in the close 
vicinities of Europa, Ganymede and Callisto. 
No dust cloud could be measured close to Io because the 
spacecraft orientation prevented the detection of dust particles 
during all fly-bys at this satellite. 

\begin{figure}
\vspace{-3mm}
\hspace{2.5cm}
\epsfxsize=0.6\hsize
\epsfbox{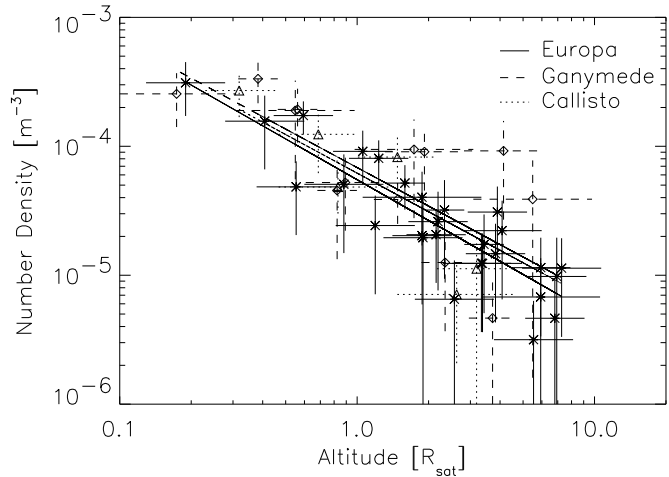}
\vspace{-7mm}
\caption[Number density of dust as a function of altitude above the surface
of Europa, Ganymede and Callisto]
{ Number density of dust as a function of altitude above the
  surface of Ganymede (data from 4 fly-bys), Europa  (8 flybs) and
  Callisto (3 fly-bys). The altitude is shown in units of
  the satellite radius $\mathrm{R_{sat}} = 1560, 2634, 2409 \,\mathrm{km}$ in the
  case of Europa, Ganymede and Callisto, respectively.
  Vertical error bars reflect statistical uncertainty due to the small
  number of impacts.  The solid lines are least squares fits to the
  measured number densities.
}
\label{dust_cloud}
\end{figure}

Analysis of the impact directions and impact speeds showed that 
the grains belonged to steady-state dust clouds surrounding the 
satellites \cite{krueger1999d,krueger2000}. The measured radial 
density profiles of the dust clouds 
(Fig.~\ref{dust_cloud}) together with detailed
modelling of the impact-ejection process implied that the particles 
had been kicked up by hypervelocity impacts of micrometeoroids 
onto the satellite's surface \cite{krivov2002b}.
The projectiles were most likely interplanetary dust particles. 

The measured mass distribution of the grains was consistent with 
such an ejection mechanism with grain sizes being 
mostly in the range 0.5 \mum\ $\leq s \leq $ 1.0 \mum. It implied that
the particle dynamics was dominated by gravitational forces, 
whereas non-gravitatonal, especially electromagnetic forces 
were negligible. Most ejected grains follow ballistic trajectories 
and fall back to the surface within minutes after they have been released.
Only a small fraction of the ejecta has sufficient energy to 
remain at high altitudes for several hours to a few days. 
 Although they eventually strike the satellite's surface, 
these short-lived but continuously replenished particles form a tenuous 
steady-state dust cloud which entirely envelopes the satellite. 
The total amount of debris contained in such a steady-state cloud is 
roughly 10 tons. 

The optical thickness of the cloud is by far too
low to be detectable with imaging techniques. Only a highly
sensitive detector of the Galileo/Ulysses type could recognize 
a sufficient number of grains to detect these clouds.  The low dust 
density is illustrated by the fact that  
only 35 cloud particles impacted the detector during 
4 fly-bys at Ganymede \cite{krueger1999d}. 

A detailed analysis of the entire Galileo dataset for the three
Galilean satellites is ongoing. One goal is to check for 
signatures of a leading-trailing asymmetry of the ejecta clouds,
which can be expected from the orbital motion of the satellite 
with respect to the field of impactors \cite{sremcevic2002}. 

The Galileo 
measurements are the first successful in-situ detection of satellite 
ejecta in the vicinity of a source moon. 
All celestial bodies 
without gaseous atmospheres (asteroids, planetary satellites of all sizes)
should be surrounded by an ejecta dust 
cloud.    Before Galileo, there were few attempts of direct in-situ
detections of ejecta close to satellites~---
most notably, near the Moon \cite{iglseder1996}.
These experiments, however, did not lead to definite results. 

\section{Dusty Jovian rings}

\label{dust_rings}

\begin{figure}
\vspace{-2.0cm}
\epsfxsize=0.9\hsize
\epsfbox{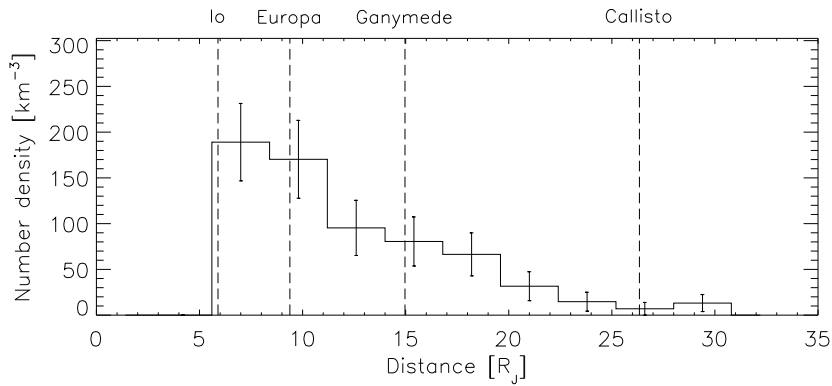}
\vspace{-12.4cm}
\caption{\label{dust_ring}
Number density of micrometer-sized dust particles between 6 and 30\RJ\ 
constructed from the Galileo measurements. The orbits of 
the Galilean satellites are indicated by vertical dashed lines.
Only data from Galileo orbits I24 to I33 are shown because earlier
orbits only partially traversed the region between Io and Europa.
Dust cloud particles identified in the close vicinity of the moons
are not shown. 
}
\end{figure}

Apart from Io dust streams (Sect.~\ref{dust_streams}) and 
circum-satellite ejecta-clouds (Sect.~\ref{dust_clouds}) the in-situ
Galileo measurements have revealed additional populations of 
Jovian dust (Tab.~\ref{tab_dustpop}): since the beginning of Galileo's orbital 
tour about Jupiter the dust detector has measured more than 400 impacts 
of mostly micrometer-sized grains widely distributed in circum-jovian 
space. Although the highest fluxes of grains occurred in the region between  
Io's and Callisto's orbit ($\sim 6$ to 26\RJ\ from Jupiter, 
\cite{gruen1998,krueger1998}, Fig.~\ref{dust_ring}) 
impacts were also detected out to 200\RJ\ and beyond.
These grains form a tenuous dust ring around Jupiter with a number 
density of $\rm \approx 2 \cdot 10^{2}\,km^{-3}$ at Europa's orbit.
The spatial locations where these grains
were detected, the impact directions and the charge signals
imply that these are actually two populations: besides a population of 
particles on prograde orbits 
about Jupiter, another population on retrograde orbits must exist 
as well \cite{thiessenhusen2000}. The grains on retrograde orbits are
 most likely 
interplanetary or interstellar grains captured by the Jovian
magnetosphere \cite{colwell1998a,colwell1998b}. Numerical models show that
a tiny fraction of the impact debris released from the surface of the
satellites by hypervelocity impacts (Sect.~\ref{dust_clouds}) is 
ejected at speeds sufficient to
escape from the satellites entirely \cite{krivov2002a} (an amount 
of $\rm 10\,g\,sec^{-1}$ has been estimated to leave Ganymede). The 
ejected material goes into orbit about Jupiter and forms a
tenuous ring of dust particles mostly on prograde orbits. This
ring extends at least from Io's orbit (5.9\RJ\ from Jupiter) out 
to Callisto's orbit (26\RJ) but the dust detections indicate
that it  continues  further out and further in (see below).

In the outer region of the Jovian system, between 50 and 300~\RJ, about 100 
dust impacts were detected. Their orbits are compatible with prograde 
and retrograde jovicentric orbits with a wide range of inclinations
\cite{krivov2002b}. The 
number densities of $\rm \sim 10\,km^{-3}$ are  more than an order of
magnitude lower than those found in the region between the Galilean 
satellites but, on the other hand, by about an order of magnitude larger 
than the interplanetary background. Sources for these grains are Jupiter's 
outer regular and irregular moons.

Indications for the existence of the ring can already be found
in earlier measurements by the Pioneer~10/11 and Ulysses
spacecraft: 12 meteoroid penetrations have been recorded with Pioneer
within 45\RJ\ (Jupiter radius, \RJ\,= 71,492\,km) 
from Jupiter \cite{humes1974} and Ulysses has recorded 9 
impacts of micrometer-sized dust grains in this spatial region. 
Two-third of the Ulysses impacts were detected at $\sim 35^{\circ}$ 
jovigraphic latitude after Jupiter fly-by. 

Between Io's orbit at 5.9\RJ\ and 
the outer extension of the gossamer ring at about 3.1\RJ,
extremely little is presently known about the dust 
environment. Although Galileo has traversed part of this region
during orbit insertion in December 1995, dust measurements were 
very patchy because the instrument had to be saved from the 
hazards of Jupiter's radiation environment. However, a few probably
micrometer-sized dust impacts were detected within 
Io's orbit \cite{gruen1996c,krueger1999a}.

Still closer to Jupiter lies the region of Jupiter's prominent 
ring system which consists of three components: the main ring, the halo
and the tenuous gossamer rings. Here, the dust densities are so large 
that dust investigations habe been performed with remote sensing techniques.
The vertical extension and density profiles of the rings imply that 
a significant fraction if not all of the dust forming the rings is
impact-ejecta derived from the inner moons Adrastea and Metis (in the 
case of the main ring), and Amalthea and Thebe (in the case of  the
gossamer rings \cite{ockert-bell1999}). These satellites orbit
Jupiter inside the ring system. The motion of the dust grains in a certain
size range contained in the gossamer ring is most probably dominated 
by the Poynting-Robertson drag force, indicating that the plasma 
density in this region is much lower than previously thought
\cite{burns1999}. 


\begin{table*}
\caption{\label{tab_dustpop}
Physical parameters of dust populations detected in-situ at Jupiter. Column 2 gives
typical particle sizes (radii) assuming spherical particles, col. 3 the mean measured 
impact speeds, col. 4 lists the
radial distance range where the particles have been detected, and col. 5 gives
derived particle number densities in space. 
}
\begin{center}
\small
\begin{tabular}{lccccc}
\hline
\multicolumn{1}{c}{Population}        & 
Particle           & 
Impact             & 
Jovicentric        &
Number             \\
                   & 
size               & 
speed              & 
distance           &
density            & 
                   \\
                   & 
($\mu \mathrm{m}$) & 
(\kms)             &
                   &
($\mathrm{m^{-3}}$)&
         \\
\multicolumn{1}{c}{(1)} &
(2)                &
(3)                &
(4)                &
(5)                \\
\hline
Stream particles   & 
$\sim 0.01^{\ast}$ &
$\leq 400^{\ast}$  &
6\RJ -- 2\,AU      & 
$10^{-3} - 10^{-8}$\\
Ejecta clouds      &
0.3 -- 1           &
$ 6 - 8 $          &
$\leq 10\,\mathrm{R_{sat}}^{\dagger}$ &
$10^{-4} - 10^{-5}$\\
Ejecta ring         &
0.6 -- 2           &
$ \sim 7 $         &
6 -- 30\RJ         &
$10^{-6} - 10^{-7}$\\
Captured particles  &
$ 0.5 - 1.5 $      &
$\sim 20 $         &
6 -- 20\RJ         &
$\sim 10^{-7}$     \\
Outskirts ring     &
1 -- 2             &
$ \sim 5 $         &
$\geq 50$ \RJ      &
$\sim 10^{-8}$     \\
\hline
\end{tabular}
\end{center}
$\ast$: derived from dynamical modelling.

$\dagger$: altitude above satellite surface.
\end{table*}

\section{Conclusions and outlook}

The Galileo dust measurements have drastically expanded our knowledge 
about dust in the Jupiter system. In fact, Jovian dust has been
studied to at least a similar extent as cosmic (i.e. non-artificial) 
dust in the Earth environment. The properties of the various Jovian dust 
populations studied in-situ with Galileo are summarised in 
Tab.~\ref{tab_dustpop}.

All Galilean satellites are sources of dust in the Jovian system. 
The Galileo measurements have for the first time demonstrated the
electromagnetic interaction of charged dust grains with a planetary 
magnetosphere. Jupiter's magnetosphere acts as a giant mass-velocity 
spectrometer for charged dust grains in space. The Io dust 
streams can be used as a potential monitor of the activity of 
Io's plume activity.

The Io dust stream particles probe the conditions in the Io plasma torus.
Since in a completely radially symmetric plasma and magnetic field 
configuration no 10\,h period should show up in the impact rate, only 
the 5\,h period should be there. The prominent modulation of the rate 
with the 10\,h period points to variations in the acceleration 
mechanism of the grains correlated with Jovian local time which are 
presently not completely understood.

In February 2004, Ulysses will approach Jupiter to 0.8 AU again.
Additional dust stream measurements with Ulysses in interplanetary space at 
high jovigraphic latitudes and for varying Jovian local times
will be beneficial to test our understanding of this new phenomenon.

The Galileo measurements of impact-generated dust clouds surrounding 
the Galilean satellites can be considered as  unique natural impact 
experiments to study the dust ejection mechanism due to hypervelocity 
impacts 
onto celestial bodies without atmospheres. They complement laboratory 
experiments in an 
astrophysically relevant environment. Although far from being perfect 
impact experiments, the Galileo results offer two extremely important
improvements over laboratory experiments: 1) the projectile 
and target materials and projectile speeds are astrophysically 
relevant, and 2) the masses and speeds of the ejecta particles can 
be determined in an important region of parameter space 
(micrometre sizes and \kms\ impact speeds). 
This is especially important in view of the Cassini 
mission. Cassini will start its exploration of the Saturnian system in 2004 
and will fly by several of Saturn's satellites during its orbital tour
about the giant planet. It will provide a unique opportunity to study the 
dust environments of many of the small Saturnian satellites. 

\begin{figure}[tb]
\parbox{0.65\hsize}
{
\vspace{-4.5cm}
\hspace{-3mm}
\epsfxsize=\hsize
\epsfbox{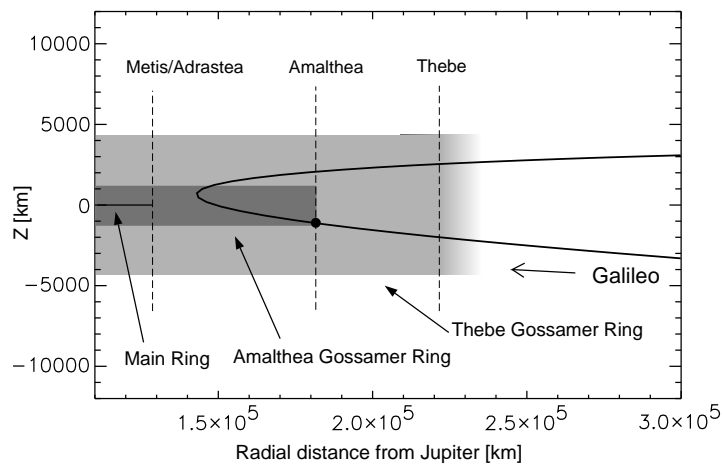}
}
\parbox{0.34\hsize}
{
\vspace{-2.7cm}
\caption{\label{A34_geometry}
Galileo trajectory during the passage of the Jovian gossamer rings in November 2002.
The position of Amalthea during closest approach is shown as a filled 
circle.
}
}
\vspace{-2.5cm}
\end{figure}

Although considered to be the archetype of an ethereal dusty planetary ring, 
the Jovian gossamer and main ring system has been relatively incompletely studied 
to date. 
The in-situ measurements of ejecta grains escaping from the circum-satellite
dust clouds and images of Jupiter's main and gossamer rings 
have demonstrated that impact-ejecta derived from 
hypervelocity impacts onto satellites is the major -- if not the only --
constituent of these dusty planetary rings. The details of the complex 
dynamics of grains over a large size range and under the various forces 
acting on the grains are as yet only poorly understood. 

In November 2002 -- during its final orbit about Jupiter -- Galileo will 
traverse the gossamer ring system and fly by 
Amalthea (Fig.~\ref{A34_geometry}). 
Detailed in-situ studies of the dust grains
in the gossamer rings will provide a better understanding of the forces 
dominating the grain dynamics in the rings (gravity, Lorentz force,
plasma drag, Poynting-Robertson drag, radiation pressure). The relative 
importance of each force varies strongly with grain size and distance from the 
planet and leads to drastically different size distributions at different 
locations along the gossamer rings and in the main ring. Investigation of 
why the Poyting-Robertson drag dominates over the other forces will lead to
a comprehensive picture of the grain dynamics in the gossamer ring, a 
neccessary
step in deriving a full picture of the dust dynamics throughout the 
Jovian magnetosphere. In-situ studies of the ring material can provide 
valuable information about the surface properties of the source moons. 
Comparative studies of ejecta from the large Galilean moons and the smaller
ones embedded in the gossamer rings will provide information about the 
ejection process over a large range in speed not accessible in the 
laboratory. Especially the close fly-by at Amalthea ($<300\,$km) 
will allow to test on a small moon the models for the impact-ejection 
process which have been developed for the much
larger Galilean satellites.

In-situ dust measurements  provide information
about the physical properties of the dust environment 
 not accessible with imaging techniques. Since all dusty 
planetary rings in our solar system are most likely dominated by 
impact-ejecta, studies of Jupiter's gossamer ring provide 
valuable information not only about the mechanism feeding this ring 
system but also about the processes that govern planetary 
rings in general. Studies of the Jovian ring with Galileo and of the 
Saturnian ring with Cassini will lead to a vastly improved understanding of
the formation and evolution of dusty planetary rings.

\vspace{0.5cm}

\noindent
{\bf Acknowledgements}

\noindent
We wish to thank the Ulysses and Galileo projects at ESA and NASA/JPL for effective
and successful mission operations. We also wish to express our gratefulness 
to our co-investigators. Without their continuous efforts over the years
the results presented here would not have been achievable.
We are also grateful to John R. Spencer and John Stansberry who provided 
thermal near-infrared data of Io.
This research has been supported by the German Bundesministerium f\"ur Bildung 
und Forschung through Deutsches Zentrum f\"ur Luft- und Raumfahrt e. V. 
(DLR, grant 50 QJ 9503 3).
Support by Deutsche Forschungsgemeinschaft (DFG)
and Max-Planck-Institut f\"ur Kernphysik (MPIK)
are also gratefully acknowledged.

\bibliography{pape,references}

\end{document}